\documentclass[usegraphicx,usenatbib]{mn2e}
\usepackage{times}
\usepackage{epsf}

\newcommand{\source}{\hbox{NGC\,4261}}
\newcommand{\sourcer}{\hbox{3C\,270}}
\newcommand{\asca}{\textit{ASCA}}
\newcommand{\chandra}{\textit{Chandra}}
\newcommand{\rosat}{\textit{ROSAT}}
\newcommand{\einstein}{\textit{Einstein}}
\newcommand{\xmm}{\textit{XMM-Newton}}
\newcommand{\hst}{\textit{HST}}
\newcommand{\vla}{\textit{VLA}}

\title[The jet and counterjet of \sourcer]
{The jet and counterjet of \sourcer\ (\source) viewed in the X-ray
with \chandra}

\author[D.M. Worrall et al.]
 {D.M. Worrall,$^1$
M. Birkinshaw,$^1$
E. O'Sullivan,$^2$
A. Zezas,$^{3,4,5}$
A. Wolter,$^6$
G. Trinchieri$^6$
\newauthor
and G. Fabbiano$^5$
\\
$^1$HH~Wills Physics Laboratory, University of Bristol, Tyndall Avenue,
Bristol BS8~1TL \\
$^2$School of Physics and Astronomy, University of Birmingham, Edgbaston,
 Birmingham B15 2TT \\
$^3$Physics Department, University of Crete, 71003 Heraklion, Greece \\
$^4$IESL, Foundation for Research and Technology--Helas, 7110 Heraklion, Greece \\
$^5$Harvard-Smithsonian Center for Astrophysics, 60 Garden Street,
 Cambridge, MA 02138, USA \\
$^6$INAF -- Osservatorio Astronomico di Brera, via Brera 28, 20121
 Milano, Italy \\
 }

\begin{document}

\label{firstpage}

\maketitle

\begin{abstract}

The radio source \sourcer, hosted by nearby elliptical galaxy \source,
is the brightest known example of counterjet X-ray emission from a
low-power radio galaxy.  We report on the X-ray emission of the jet
and counterjet from 130~ks of \chandra\ data.  We argue that the X-ray
emission is synchrotron radiation and that the internal properties of
the jet and counterjet are remarkably similar. We find a smooth
connection in X-ray hardness and X-ray to radio ratio between the jet
and one of the X-ray components within the core spectrum.  We observe
wedge-like depressions in diffuse X-ray surface brightness surrounding
the jets, and interpret them as regions where an aged population of
electrons provides pressure to balance the interstellar medium of
\source.  About 20~per cent of the mass of the interstellar medium has
been displaced by the radio source.  Treating \sourcer\ as a twin-jet
system, we find an interesting agreement between the ratio of
jet-to-counterjet length in X-rays and that expected if X-rays are
observed over the distance that an outflow from the core would have
travelled in $\sim 6 \times 10^4$ yr. X-ray synchrotron loss times are
shorter than this, and we suggest that most particle acceleration
arises as a result of turbulence and dissipation in a stratified flow.
We speculate that an episode of activity in the central engine
beginning $\sim 6 \times 10^4$~yr ago has led to an increased velocity
shear.  This has enhanced the ability of the jet plasma to accelerate
electrons to X-ray-synchrotron-emitting energies, forming the X-ray
jet and counterjet that we see today.

\end{abstract}

\begin{keywords}
galaxies: active -- 
galaxies: individual: (\sourcer, \source)  --
galaxies: jets -- 
X-rays: galaxies
\end{keywords}

\section{Introduction}
\label{sec:intro}%1

Low-power radio galaxies appear to be tuned such that their average
jet kinetic powers, which are dictated by the rate of accretion onto a
central supermassive black hole, provide the heating and
momentum input required to limit star formation in their host galaxies
by keeping the gas hot \citep[e.g.,][]{granato, kawata, best,
schawinski}.  This adds to the interest in studying jets and their
lifecycles, and one issue of importance in the study of this feedback
process is how the kinetic power of individual sources varies with
time.  Historically the radio emission has been used as a probe.
However, the loss lifetimes of electrons responsible for the radiation
exceed flow times along many resolvable jet structures, and thus the
radiation that is observed holds only a time-averaged trace
of changes to the central power output and local environmental effects
along the jet.  X-ray studies of low-power radio galaxies came into
fruition with \chandra, whose arcsec-scale resolution
\citep{weisskopf} supports the common detection of resolved X-ray
synchrotron emission from their kpc-scale jets \citep*{worrall01}.
For X-ray emission the electron energy-loss lifetimes are short
compared with light travel times over the structures that are
detected, and a focus of much current work is therefore the use of
X-ray emission as a probe of distributed particle acceleration
\citep[see][for a review]{worrall}.

Most resolved X-ray jets in low-power radio sources correspond to the
brighter radio jet only, indicating that the X-rays are not detected without
the assistance of relativistic boosting
\citep*[e.g.,][]{hard-66b, pesce-3c371, marsh-hetgsm87, harris-3c129,
birk-pks0521, evans-n6251, worrall-n315, samb-s52007}. While not
unexpected, such effects are an unwelcome complication in the
interpretation of measurements.  A further
complication arises where radio sources show gross asymmetries and
bends that must largely result from environments that are different
on each side of the nucleus.  In order to alleviate these effects, we
have chosen to study the nearby radio source \sourcer, whose twin jets
lie close to the plane of the sky and whose symmetry is largely
intact.  In this paper we report a deep X-ray observation of its jet
and counterjet.

\begin{figure}
\centering
\includegraphics[width=2.5truein]{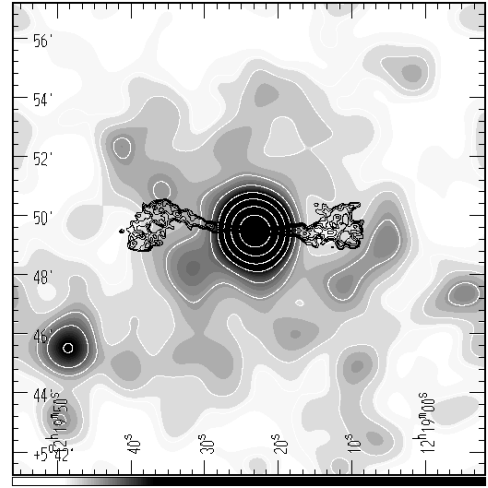}
\caption{
Large-scale radio and X-ray structure of \source.  The black contours are of a
5~GHz VLA radio map \citep{birkd} made with a 
$11.5 \times 10$ arcsec$^2$ beam; the 12 contours are logarithmically spaced
between 12 and 220 mJy beam$^{-1}$.
The \rosat\ PSPC
0.4--2~keV image is smoothed with a Gaussian of
$\sigma = 30$ arcsec. Background has been subtracted and the image is corrected
for exposure and vignetting using the {\sc  exas} software \citep{snowden}.
White X-ray contours, in units of $10^{-3}$ counts arcmin$^{-2}$ s$^{-1}$, are
1.1,1.9,3.4,6.0,11.0,19.0,34.0.
}
\label{fig:xrbw}
\end{figure}

3C\,270 is hosted by the nearby elliptical galaxy \source\ \citep[$z=
0.00746$,][]{trager}.  The galaxy's  major-axis extent to
25 B mag arcsec$^{-2}$ is $244\pm 12$ arcsec at position angle
$\sim 160^\circ$,
and the ellipticity is  0.11 \citep[][taken from
http://nedwww.ipac.caltech.edu/]{devauc}.  
The radio emission extends roughly 520 arcsec at a position angle
$\sim 88^\circ$
\citep{birkd}, and
is symmetrical on the large scale, as seen in Figure~\ref{fig:xrbw}
where the black contours of radio emission show the brightest parts of
the twin lobes.  The jet radio emission to the west is somewhat
brighter than the counterjet to the east, but the overall morphology
suggests that the radio emission is not highly affected by
relativistic beaming.  Indeed, \citet{piner} have used the apparent
speed and the jet-to-counterjet brightness ratio from VLBI
images to deduce a jet speed of $(0.46 \pm
0.02)c$ and an inclination angle of $\theta = 63^\circ \pm 3^\circ$.

X-ray emission from \source\ was first detected using \einstein\
\citep{fberg}, but the improved resolution of
\rosat\ was required for component separation.
\citet{wb} used the PSPC to separate an unresolved nuclear component
from resolved emission both spatially and spectrally,
attributing the latter to the thermal atmosphere
of the source (see also Fig.~\ref{fig:xrbw}).
Further investigation of the thermal component
with \rosat\ and \asca\ \citep{davis, osmond, finjones} found that the gas
temperature increased with distance from \source, with a group
atmosphere taking over from interstellar gas at
the outer extremities of the radio source.
Gas density and temperature profiles assuming spherical symmetry
and using the \chandra\ data discussed here appear in
\citet{hump}, and see \cite{newpaper} for
further discussion of the galaxy and group gas properties. 

Improved measurements of the X-ray core using \xmm\ and an earlier
32-ks \chandra\ observation \citep{sambruna, gliozzi, chiaberge,
donato, zezas} revealed that part of the nuclear emission suffers from
high intrinsic absorption, with $N_{\rm H}$ several times
$10^{22}$ cm$^{-2}$ (see also Section~\ref{sec:xnucleus}) contrary to
the common trend towards low intrinsic absorption in radio galaxies
whose bolometric luminosities are small compared with their Eddington
limits \citep{evans-cores}.

Resolved X-ray emission from both the jet and counterjet were first
seen in the earlier \chandra\ observation \citep{chiaberge, zezas},
showing \sourcer\ to be the brightest known example of counterjet
X-ray emission from a low-power radio galaxy.  The radio jets close to
the nucleus on either side are largely undisturbed, and \sourcer\ is thus an
excellent source for addressing jet physics.  In this paper we
concentrate on the resolved X-ray jet and counterjet, and their
connection with the unresolved nucleus and their surroundings.  Based
on our results, we suggest that X-ray synchrotron emission in
low-power jets may be used not only to address issues of particle
acceleration but also to probe time-variable power released by an
active-galaxy nucleus.

We adopt $H_0 = 70$~km~s$^{-1}$~Mpc$^{-1}$.
1~arcmin corresponds to 9.2~kpc at the distance of
NGC~4261.

\section{\chandra\ X-ray observations and analysis}
\label{sec:obs}%2

We observed \source\ in FAINT data mode with the Advanced CCD Imaging
Spectrometer (ACIS) on board \chandra\ on 2008 February 12 (OBSID
9569, sequence 600710).  Details of the instrument and its modes of
operation can be found in the \chandra\ Proposers' Observatory
Guide\footnote{ http://cxc.harvard.edu/proposer}.  Results presented
here use {\sc ciao v4.1.2} and the {\sc caldb v4.1.3} calibration
database.  We re-calibrated the data, with random
pixelization removed and bad pixels masked, following the software
`threads' from the \chandra\ X-ray Center (CXC)\footnote{
http://cxc.harvard.edu/ciao}, to make a new level~2 events file.  Only
events with grades 0,2,3,4,6 were used, as recommended. After removal of time
intervals when the background deviated more than $3\sigma$
above the average value, the
calibrated dataset from 2008 has an observation duration of 100.538~ks.

We had earlier made a shorter observation of \source\ with \chandra\
on 2000 May 6 (OBSID 834, sequence 700139). In 2000 the source was
observed with the S3 CCD chip in 512-row subarray and in VFAINT mode,
giving a 4 by 8 arcmin field of view.  The subarray was used to reduce
the readout time to 1.82~s and so decrease the incidence of multiple
events arriving from the nucleus within the frame-transfer time (so
called pile-up).  In the event, we found that pile-up was less than
3\% and spectral-fitting results were unaffected \citep{zezas}, and so
for our 2008 observations we exposed the whole 8 by 8 arcmin field of
the S3 chip, with a frame time of 3.24~s.  
Some surrounding CCDs were also exposed, but only data from the S3 chip
are considered in this paper.  The jet was advantageously misaligned
with the read-out direction in both observations.  To take advantage
of the latest calibrations we have also made new level~2 events files
for the earlier data, with and without VFAINT cleaning to help remove
particle background.  The calibrated dataset from 2000 has an
observation duration of 32.249~ks.

We made a merged events file of the two observations for image-display
purposes.  For spectral fitting we have extracted data files and
calibrations for the two observations separately, and we normally fit
them jointly to models.  For spectral extraction we use the {\sc ciao}
task {\sc specextract}, except for the core spectrum where we use {\sc
psextract} which is more suitable for a point source, followed by {\sc
mkacisrmf} and {\sc mkarf} to apply the latest calibrations.  For the
2000 data, the core spectrum is extracted from the file where VFAINT
cleaning has not been applied, since this cleaning removes some events
that are likely to be real for a bright point source.  Spectra are
binned to a minimum of 25 counts per bin for $\chi^2$ fitting with
Gaussian errors using {\sc xspec}.  All spectral fitting was performed
over the energy range 0.3--10~keV, and includes absorption along the
line of sight in our Galaxy assuming a column density of $N_{\rm H} =
1.58 \times 10^{20}$ cm$^{-2}$ (from the {\sc colden} program provided
by the CXC, using data of \citet{dlock90}).

Parameter values are quoted with errors corresponding to 90\%
confidence for one interesting parameter ($\chi^2_{\rm min} + 2.7$,
with all other interesting parameters allowed to vary), unless
otherwise stated.  For non-thermal components, spectral index,
$\alpha$, is defined in the sense that the flux density is
proportional to $\nu^{-\alpha}$: the photon spectral index is $\alpha
+ 1$, and the number power-law spectral index of radiating electrons
is $p = 2\alpha + 1$.

\section{Results}
\label{sec:results}%2

\subsection{Imaging}
\label{sec:imaging}%2

\begin{figure}
\centering
\includegraphics[width=3.5truein]{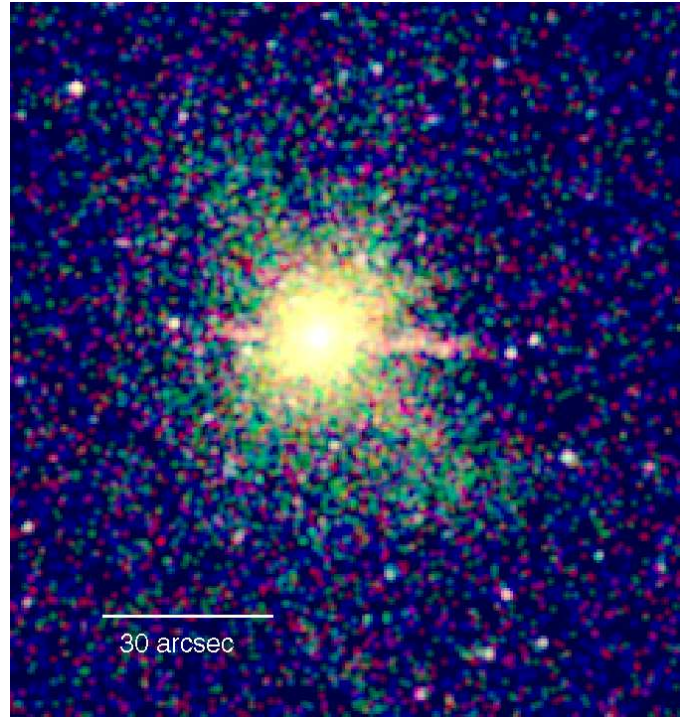}
\caption{
False-colour \chandra\ image from the combined \chandra\
observations.
Red is 0.3-0.8~keV, green is 0.8-1.2~keV, blue is 1.2-5~keV.
The images have each been smoothed with a Gaussian of $\sigma =
1$~arcsec before being combined.
}
\label{fig:ximage}
\end{figure}

\begin{figure}
\centering
\includegraphics[width=3.5truein]{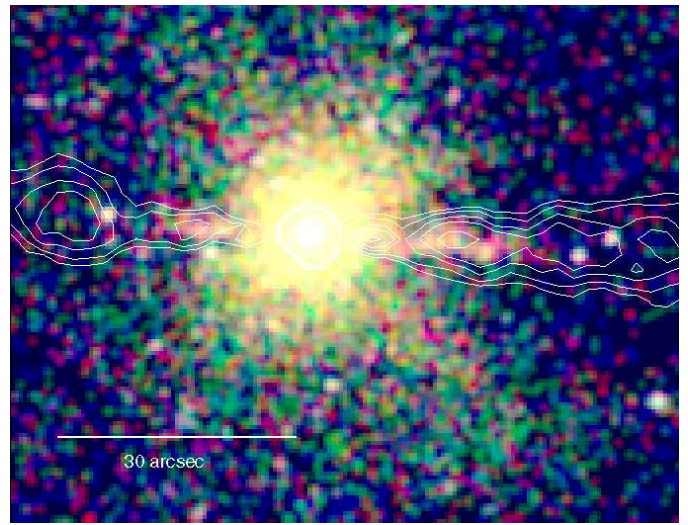}
\caption{
Zoomed version of Fig~\ref{fig:ximage}
overlayed with contours from a
5~GHz \vla\ radio map \citep{birkd} made with a 
$2.3 \times 1.8$ arcsec$^2$ beam; contours at 0.7, 1, 1.5, 2, 3 mJy beam$^{-1}$.
}
\label{fig:xrimage}
\end{figure}

Our 100-ks \chandra\ exposure reveals more clearly the resolved jet
and counterjet, whose presence was noted in the earlier shorter
exposure by \citet{chiaberge} and \citet{zezas}.
From inspection, the jet and counterjet substructures are the
same in the two observations and there is no obvious sign of
point-like variability.  We have therefore combined the data from the two
observations. 
However, our ability to assess variability
is limited by the low number of net counts and the strong change in low-energy quantum
efficiency of the detector between the observations (see
Section~\ref{sec:xjet} for more information).
Figure~\ref{fig:ximage} shows a false-colour representation of the
data.  The jet (to the W) and
counterjet (to the E) are clearly detected out to 31.7 and 20.2
arcsec, respectively. Both are whiter in colour (i.e., emit relatively
more flux above $\sim 1$~keV) than the galaxy emission, which is asymmetric and
flares, particularly to the SW.  In both the jet and counterjet the
emission appears to be most red (i.e., has the steepest spectrum)
furthest from the nucleus (see also Section \ref{sec:xjet}). The
colour representation shows two point sources, 34.2 and 38
arcsec from the nucleus to the W, and one that is 25.9 arcsec to the
E.  All three are significantly bluer in colour than either the jet or
the diffuse emission in the galaxy, and we do not believe them to be jet
features.  The jet and counterjet X-ray emission correspond to the
regions where the radio emission is brightest, as shown in
Figure~\ref{fig:xrimage}.

\begin{figure*}
\centering
\includegraphics[width=6.5truein]{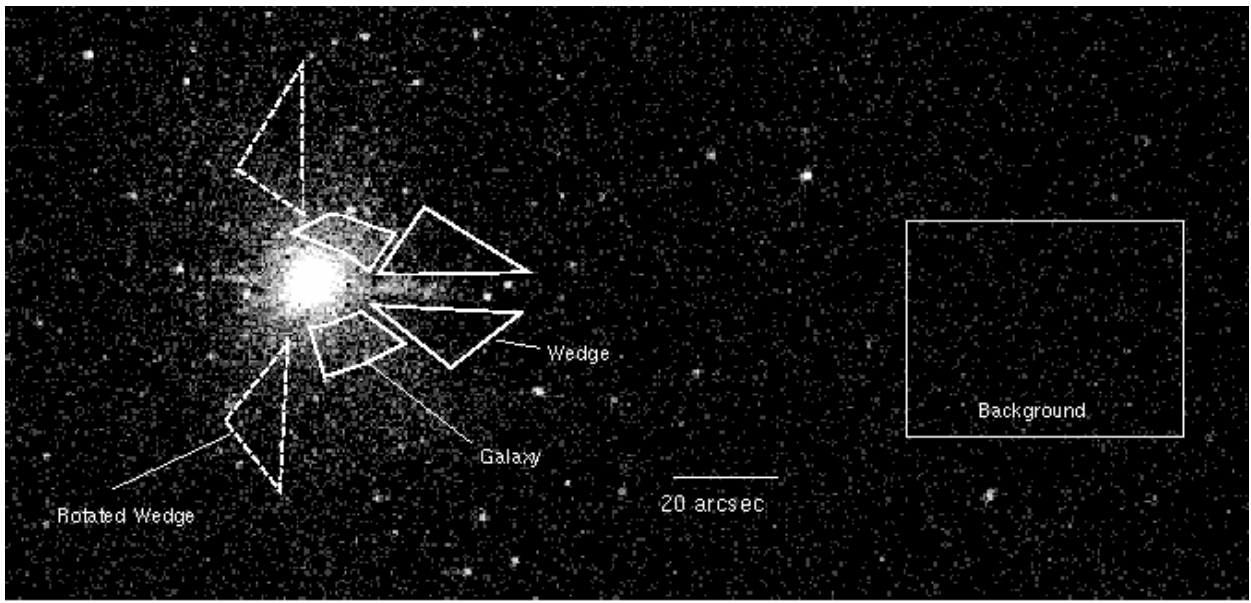}
\caption{
0.3--5~keV unsmoothed X-ray image from the combined \chandra\ observations
in $0.492$ arcsec pixels overlayed with shapes marking regions used in
spectral extraction.  The rectangle is the background region for the
shapes close to the source.  The dashed regions shows the wedges
rotated by 90 degrees.
}
\label{fig:gasregions}
\end{figure*}

Particularly striking in the new data are wedge-like deficits of
diffuse X-ray emission around the jet (see also
Fig.~\ref{fig:gasregions}), and to a lesser extent around the
counterjet.  These are by far the most prominent structural
features in the galaxy-scale gas distribution.  Their lifetime would
be of order the sound crossing time, $\la 20$ Myr, unless they are
maintained.  This maintenance could be provided by the radio source,
whose overall lifetime is likely to be of order 20 Myr based on
forward expansion at several times the speed of sound. The
correspondence of these lifetimes and of the wedge geometry with the
jet direction would be coincidences if the wedges were temporary
features associated with the galaxy encounter suggested by the faint,
and almost certainly older, NW tidal arm reported by \citet{tal}.  The
results therefore suggest that the radio plasma has displaced,
transversely, some of the X-ray-emitting atmosphere.

Emission from galaxy gas is prominent in the images, and
even in the outskirts of Figure~\ref{fig:gasregions} the X-ray
emission is largely from gas on the group scale.  Many of the point
sources in the image are low mass X-ray binaries (LMXBs), and these
are discussed in \cite{newpaper2}.

\subsection{The gas around the resolved X-ray jet}
\label{sec:gasspectra}

We have compared the emission-weighted temperature of gas in the
two wedge-like regions around the jet that are deficient in X-rays
(hereafter called `wedge gas') with
that of nearby galaxy gas (hereafter called `galaxy gas').
The regions are outlined by solid lines in
Figure~\ref{fig:gasregions}, where the triangles sample the wedges and
the irregular shapes sample the galaxy gas.
In both cases the rectangular region shown in the figure is taken as
background: it is within the region where group gas of
$kT \sim 1.3$~keV has been
measured \citep{finjones, osmond, hump}.

We have fitted the galaxy gas with a thermal (APEC) model.  A
contribution from faint unresolved LMXBs is expected, and so we have
accounted for this by including a thermal bremsstrahlung component,
since that has been found to give a good fit to the cumulative
spectrum of the LMXBs of an elliptical galaxy \citep*{sarazin, irwin},
and has the advantage of being described by just the two parameters of
temperature and normalization.  The bremsstrahlung temperature is
poorly constrained, and following \citet{hump} we fix it to $kT
=7.3$~keV, the well-constrained best fit for 15 nearby
early-type galaxies as found by \citet{irwin}.  The gas
abundances are poorly constrained, and so these have been fixed at
solar.  Our combined model gives $\chi^2 = 42.8$ for 55 degrees of
freedom, as compared with $\chi^2 = 64$ for no LMXB contribution.  The
LMXB bremsstrahlung contribution is $8\pm 3$\% of the total 0.3--2~keV
luminosity.  The galaxy gas has $kT = 0.63^{+0.03}_{-0.02}$~keV, which
agrees with earlier results from \rosat, \asca, \xmm, and \chandra\
for roughly this distance from the nucleus \citep{wb, finjones,
gliozzi, zezas, hump}.  The same model gives a good fit to the wedge
gas.  The only significance difference (bearing in mind that errors
are quoted at 90\% confidence) is that the gas temperature is higher
in the wedge region, at $kT = 0.78 \pm 0.09$~keV ($\chi^2 = 16.6$ for
16 degrees of freedom for the best fit).

In order to test if the special location next to the jet is
responsible for the higher temperature in the wedge than in the galaxy
gas with which the wedges are partially in contact, we
rotated the northern wedge by 90~degrees and the southern by
-90~degrees, keeping each vertex the same distance from the nucleus as
before (see regions outlined by dashed lines in
Fig.~\ref{fig:gasregions}).  While these new regions sample
gas at a similar distance from the core than that in the wedges, they
miss the brighter plumes of galaxy atmosphere.  We found that the
rotated wedges gave a similar temperature to that in the wedges
surrounding the jet: $kT = 0.81 \pm 0.05$~keV ($\chi^2 = 19.8$ for 21
degrees of freedom for the best fit).  The emission in the wedge
regions is indeed weaker (in the energy band 0.3--2 keV there are
$261\pm19$ net counts in the wedges as compared with $438\pm23$ net
counts in the rotated wedges).  However, we conclude that the gas
surrounding the jet is at a typical temperature for its distance from
the nucleus.  The displacement of gas in the wedge regions is
therefore not accompanied by significant local heating. This is
consistent with the absence of edge-brightenings that would suggest
supersonic transverse expansion of the cavities, and so implies that
these cavities are close to pressure equilibrium with the external
gas. The pressure in the cavities can be supplied by the particles and
fields in a diffuse radio-emitting plasma that lies around the radio
jets but does not appear on the high-frequency and high-resolution
radio contour images in Figures~\ref{fig:xrbw} and \ref{fig:xrimage}
\citep[but see the map in][]{newpaper}. The X-ray spectrum of the wedge regions is
not inconsistent with this idea: fits incorporating some 0.63~keV gas
associated with the inner galaxy atmosphere and 0.9~keV gas associated
with the outer atmosphere are permitted in the rotated wedge regions,
and the wedge region spectrum can be fitted adequately just with $\sim
0.9$~keV gas with the same normalization as in the rotated wedge. 
More precise X-ray spectra would be needed to test this model further.

\subsection{The X-ray jet and counterjet}
\label{sec:xjet}

We have measured the jet spectrum extracted from a region of position
angle 265.5 degrees, width 4.7 arcsec, and length 22.9 arcsec starting
at 8.8 arcsec from the nucleus.  The counterjet region is defined as
being at position angle 85.5 degrees, width 4.7 arcsec, and length
11.5 arcsec, again starting 8.8 arcsec from the nucleus.  With the
wedge-shaped regions shown on Figure~\ref{fig:gasregions} to measure
the background, we found that the jet spectrum is fitted by a power
law plus a component of thermal emission at $kT \sim 0.7$~keV.  The
need for the thermal component was clear in the shape of the
residuals, and $\chi^2$ reduced by 13 when this component was added.
This is as expected since the jet extraction region extends closer to
the core than the wedge-shaped regions.  When we instead sampled the
background from regions the same size and shape as the jet and on
either side of it, we found a good fit to a power-law component alone.
The absorption gave a good fit to the Galactic value, and this was
fixed.  For the jet, which contains 410 net counts (0.3--5~keV) we
found for the power law $\alpha = 1.22 \pm 0.22$ and, over the band
0.3--5 keV, an unabsorbed flux of $(1.7\pm0.2) \times 10^{-14}$ ergs
cm$^{-2}$ s$^{-1}$ and a luminosity of $(2.1\pm0.3) \times 10^{39}$
ergs s$^{-1}$.  For the counterjet we used a background region to the
north only, since a bluer point-like source lies to the south.  The
counterjet contains only 149 net counts (0.3--5~keV) and the spectral
index is poorly constrained. We found $\alpha = 0.4^{+0.8}_{-1.1}$
and, over the band 0.3--5 keV, an unabsorbed flux of $(6.6\pm0.8)
\times 10^{-15}$ ergs cm$^{-2}$ s$^{-1}$ and a luminosity of
$(8.2\pm1.0) \times 10^{38}$ ergs s$^{-1}$.

\begin{figure}
\centering
\includegraphics[width=3.0truein]{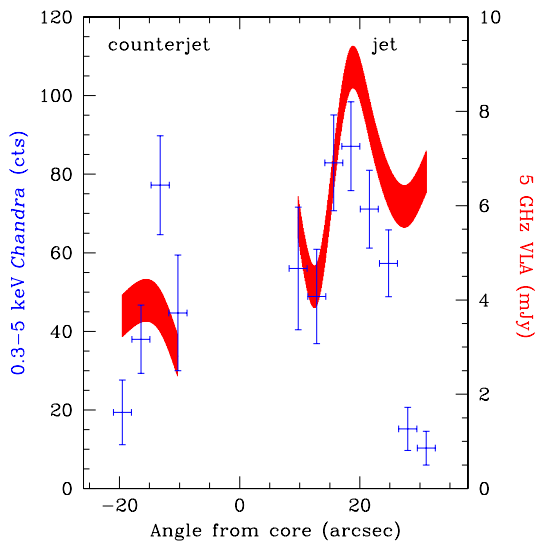}
\caption{
Profile of X-ray and radio emission in the jet and counterjet where
X-ray emission is detected and spatially separated from other
components.
The X-ray counts are shown as blue data points with $\pm1\sigma$~errors,
corresponding to values on the left-hand axis.
The radio is measured from the same regions and measurements
(corresponding
to values on the right-hand axis) are connected
with a red band whose thickness corresponds to $\pm1\sigma$~errors.
}
\label{fig:xrprofile}
\end{figure}

In the jet region, the 5-GHz radio and 1-keV X-ray flux densities are
$50.3\pm 1.3$~mJy and $2.6 \pm 0.2$~nJy, respectively ($1\sigma$
errors), giving a ratio of $(1.9 \pm 0.2) \times 10^4$.  The
equivalent values for the counterjet are $13.8\pm 0.9$ mJy and $0.8
\pm 0.3$~nJy, with a similar ratio of $(1.7 \pm 0.6) \times 10^4$.
Figure~\ref{fig:xrprofile} shows the profiles of X-ray and radio
emission in the jet and counterjet in the region where X-ray emission
is detected and spatially separated from nuclear and galaxy emission.
The subtracted background was sampled locally to each X-ray data point.
There is no significant variability in the X-ray emission 
on the angular scale of
these data points between the two observations, which were roughly
8~yr apart, and which are combined to make this plot. However, due to the low
numbers of counts we cannot rule out variability of
a factor of 2 or less within regions of area 0.3~kpc$^2$. The
plot is dominated by the 2008 data.

Spectral extraction regions for both the jet and counterjet begin
where these features are clearly separated from the point spread
function (PSF) of the bright
core.  If we assume that the jet and counterjet extend into the
nucleus with their average measured surface brightness, their estimated 0.3--5 keV
total luminosities and $1\sigma$ uncertainties are $(2.9\pm0.2)\times 10^{39}$ and
$(1.4\pm0.1)\times 10^{39}$ ergs s$^{-1}$, respectively.
Perhaps of more importance is the jet-to-counterjet intensity ratio in
the radio and X-ray.
For this we should choose identical regions on each side
of the nucleus, and so we adopt that of the counterjet and its
mirror-image superimposed on the jet.
The jet-to-counterjet intensity ratios in the radio and X-ray are then $1.8\pm0.1$ and
$1.6\pm0.3$ ($1\sigma$ errors), respectively.
There is some indication from the radio data that the
jet-to-counterjet ratio
decreases with distance from the core, as might be expected from a
slowing flow.  For example, regions of
X-ray counterjet length an X-ray-jet's length away give a ratio of
$1.4\pm0.1$.  However, our
current radio data are not of sufficient quality to explore such issues in depth.

\begin{figure}
\centering
\includegraphics[width=3.0truein]{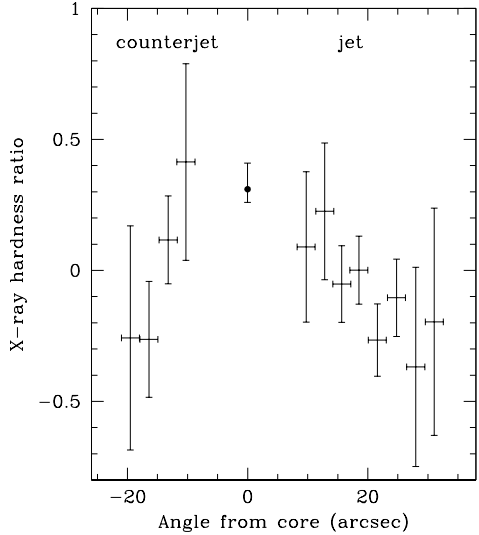}
\caption{
Profile of X-ray hardness ratio,
defined as $(H-S)/(H+S)$ where
$H$ is net counts in the band 1--5~keV, and $S$ is net counts in the
band 0.3--1~keV.
Errors are $1\sigma$.  The crosses show values for the jet and
counterjet,
using neighbouring background regions as described in Section~\ref{sec:xjet}.
The circle corresponds to component PL2 in the core spectrum, extracted
as described in Section~\ref{sec:xnucleus}.
}
\label{fig:hardness}
\end{figure}

An X-ray hardness-ratio plot (Fig.~\ref{fig:hardness})\footnote{
Calculations using the Bayesian method of \citet{park}
(http://hea-www.harvard.edu/AstroStat/BEHR/) give essentially the same
results.}  shows a trend for the X-ray spectrum to get steeper away
from the nucleus, but given the statistics the result is not highly
significant.

\subsection{The X-ray nucleus}
\label{sec:xnucleus}

We extracted the spectrum of the nucleus from a circle of
radius 1.23 arcsec.  We used background either from 
a source-centered annulus of radii $2.8$ and $8.8$ arcsec \citep[as
in][]{zezas}
or the rectangular region situated in the group gas. However, none of the
interesting parameters in the fits was affected by the choice of
background region, and results presented here use only the former region.

In agreement with the results of \citet{zezas} for the earlier,
shorter, \chandra\ exposure, we find that in the 2008 data
the spectrum requires three components, and our best fit gives a
$\chi^2$ of 114 for 107 degrees of freedom.
The first component is a contribution from galaxy gas.  The overall fit is
insensitive to the abundances, and we set them to solar, with results giving
$kT = 0.61\pm 0.03$~keV.

The second component is power-law emission of large column density
(component PL1).  This was found also by others in the earlier
\chandra\ or XMM-Newton observations \citep{sambruna, gliozzi,
chiaberge, donato}, and its origin is controversially associated
either with the accretion flow or the base of the jet
\citep[see][]{zezas}.  For this heavily absorbed component we find
$\alpha_1 = 0.5^{+0.5}_{-0.3}$, $N_{\rm H_1} = 6.6^{+2.8}_{-1.9}
\times 10^{22}$~cm$^{-2}$, and an unabsorbed 1~keV flux density of
$S_{1_{1~{\rm keV}}} = 89^{+110}_{-39}$nJy.

The third is a power law of lower absorption (component PL2).
\citet{zezas} pointed
out that the UV and optical continuum from the nucleus seen in \hst\
data is more likely to be related to this less-absorbed power-law
component than to the highly absorbed component as suggested by
\citet{chiaberge}.  We find $\alpha_2 = 1.1^{+1.9}_{-1.0}$, $N_{\rm
H_2} = 1.1^{+2.6}_{-1.1} \times 10^{21}$~cm$^{-2}$, and an unabsorbed
1~keV flux density of $S_{2_{1~{\rm keV}}} = 12^{+16}_{-5}$nJy.  While the
spectral parameters for this soft component are poorly constrained, a
fit with only one absorbed power law and a thermal component is
unacceptable ($\chi^2 = 158$ for 109 degrees of freedom).  As noted by
\citet{zezas}, a partial covering model where emission from the
nucleus is seen through patchy absorption is an alternative for the
combination of PL1 and PL2.  However, a partial covering model
implies that the emission region is extended on the scale
size of the absorbing gas, and it is reasonable to associate the most
obscured region with a region of scale-size comparable to the
accretion disc and separate it from less absorbed power-law emission.

\begin{figure}
\centering
\includegraphics[width=3.0truein]{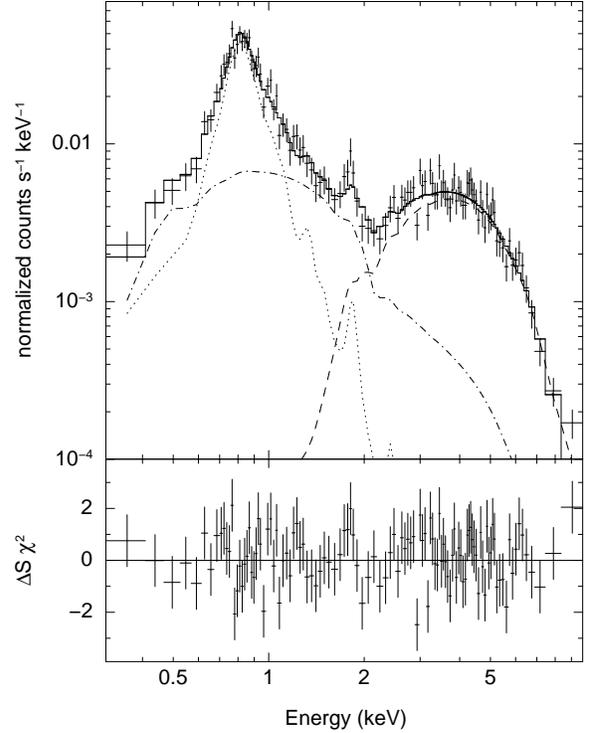}
\caption{
Core spectrum from the 2008 \chandra\ observation and best-fit model,
with the lower panel showing the residuals as their contribution to
$\chi^2$.  The dashed line is the heavily absorbed power law, the
dashed-dotted line is the less absorbed power law, and the
thermal contribution is shown as dotted.  See text for details.
}
\label{fig:corespec}
\end{figure}

The parameters for this 3-component model (two absorbed power
laws and thermal gas, see Fig.~\ref{fig:corespec}) agree within
uncertainties with values from the shorter 2000 observation published
by \citet{zezas}.  Weak evidence for a narrow Fe line has been
presented by
\citet{sambruna}, based on \xmm, and \citet{zezas}, from the
2000 \chandra\ observation.  The 2008 observation does not strengthen
this case.  There is only weak evidence for a
line around 6.4~keV in the residuals between the data and the
3-component model (Fig.~\ref{fig:corespec}).  Formally $\chi^2$ reduces
by just 3 if a narrow line of energy $6.2\pm 0.2$~keV
(equivalent width $\sim 131$~eV) is included.

A combined fit to the 2008 and 2000 \chandra\ data gives similar
parameter values to those above, with a fitting statistic of
$\chi^2 = 180$ for 159 degrees of freedom.  Freeing up the relative
normalizations of the power-law components between epochs yields no
significant improvement in fit ($|\Delta \chi^2| < 1$ for each), and
so there is no evidence for variability in either power-law
component.  In the remaining discussion, when referring to core
components we will adopt the parameter values from the combined fit:
$kT = 0.61^{+0.02}_{-0.03}$~keV, $\alpha_1 = 0.56^{+0.44}_{-0.34}$,
$N_{\rm H_1} = 7.8^{+2.2}_{-1.9} \times 10^{22}$~cm$^{-2}$,
$S_{1_{1~{\rm keV}}} = 96^{+103}_{-40}$nJy, $\alpha_2 =
0.5^{+1.0}_{-0.4}$, $N_{\rm H_2} = 2^{+14}_{-2} \times
10^{20}$~cm$^{-2}$, and $S_{2_{1~{\rm keV}}} = 8.5^{+6.1}_{-1.3}$nJy.
The 0.3-10~keV intrinsic luminosities for the hard (PL1) and soft
(PL2) power-law components are $1.4 \times 10^{41}$ and $1.4 \times
10^{40}$ ergs s$^{-1}$, respectively.  Note that PL2 contains
more than three times the total X-ray luminosity of the resolved jet
and counterjet even if they are assumed to extend to pc-scale
distances from the nucleus with constant surface brightness.

The core spectrum has been extracted from a circle of radius
190~pc, and it is possible that some of PL2 is contributed by
unresolved LMXBs.  If the ratio of LMXB X-ray emission to that of the
gas is the same as that at kpc distances from the nucleus, based on
Section~\ref{sec:gasspectra} we would expect $8\pm 3$\% of the
luminosity at 0.3-2~keV (discounting that in PL1) to be from LMXBs,
and our model fitting would be folding this into PL2.  However, since
we measure PL2 to contribute $39\pm 4$\% of the 0.3--2 keV luminosity
(again discounting PL1), we consider it unlikely that LMXBs dominate
the soft non-gaseous emission.  This conclusion is also consistent
with the luminosity function of \source's LMXBs presented by
\citet{giordano}.  Instead we focus on the evidence that PL2 is jet
related.

We note that PL2 is detected in the \chandra\ data but is not
separated from thermal emission in the larger PSF of \xmm. 
A nominal extraction circle for point sources in \xmm\ is
$30$ arcsec in radius, and we can see the issue with reference to the
spectral components in Fig.~\ref{fig:corespec} by recognizing that
an \xmm\ extraction would be 
like scaling the dotted line of the thermal emission
up by a factor of $\sim 600$ relative to the dot-dashed line of PL2.
This is a common difficulty for detecting
relatively weak soft AGN components with \xmm.
For a similar reason, the early observations with the still larger
PSFs of the \rosat\ PSPC and \asca\ did not find that the power-law emission
(needed in addition to thermal gas) contains a component that
is heavily absorbed.

\begin{figure}
\centering
\includegraphics[width=3.0truein]{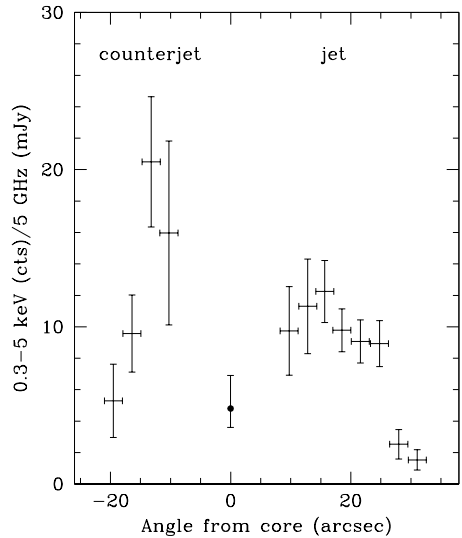}
\caption{
Profile of the ratio of 0.3--5~keV X-ray counts to 5~GHz flux density.
The data for the jet and counterjet are from Fig.~\ref{fig:xrprofile}.
The circle corresponds to X-ray emission
from component PL2 in the core spectrum (see
Section~\ref{sec:xnucleus})
divided by the radio emission from the pc-scale jet and counterjet
that are resolved with VLBI.
}
\label{fig:xrratio}
\end{figure}

VLBI mapping of \sourcer\ has found a parsec-scale radio jet and
counterjet that are well aligned with their kpc-scale counterparts and
straddle an unresolved core \citep{jwehrle}.  The core has an inverted
spectrum and varies in intensity by 30\% or more, whereas the pc-scale
jet emission is optically thin \citep{jones, piner}.  In observations
at 5~GHz taken in 1999 the core had a 5-GHz flux density of $\sim
80$~mJy while the pc-scale jet and counterjet emission totalled $\sim
300$~mJy \citep{jones}.  In the X-ray, PL1 is ten times more
luminous than
PL2 but more highly absorbed.  It is logical to assume that it arises from
a region no larger than the unresolved radio core measured with VLBI.
If we then take PL2 as primarily arising from the parsec-scale
jet and counterjet, the ratio of radio to X-ray flux density is $(3.5
^{+0.6}_{-2.5}) \times 10^4$, in agreement with the value of $\sim 1.8 \times
10^4$ for the kpc-scale structures (Section~\ref{sec:xjet}).
The validity of such a comparison rests on neither radio nor X-ray
fluxes having varied by a large factor between measurements taken some
years apart.  While we cannot be sure of this, particularly since the
separation between observations is comparable with the dynamical timescale
of the component, we note that 
no X-ray variability is measured in PL2 over 8 years.
In Figure~\ref{fig:xrratio} we show the data for the jet and counterjet
of Figure~\ref{fig:xrprofile} in the form of a ratio of X-ray to radio
emission, and have added a data-point at the core which uses X-ray
counts from PL2 and radio emission from the pc-scale jet and
counterjet.  Given that 
PL2 is roughly 40 times brighter than
an average jet and counterjet data point, it is remarkable that
its ratio of X-ray to radio emission agrees as well as it does with
the ratio for the extended emission.
This supports the idea that PL2 arises from the
pc-scale jet and counterjet through a mechanism similar to that operating
on kpc scales.

We have estimated the counts and hardness ratio over 0.3 -- 5 keV for
component PL2.  To do this we adopted the best-fitting model parameter
values and used the {\sc fakeit} task in {\sc xspec} with appropriate
response files and exposures to simulate the counts in the bands
0.3--1~keV and 1--5~keV for the 2008 and 2000 observations, which we
then combined.  Uncertainties were estimated through further
simulations that took into account the uncertainties in the model
parameters in the spectral fit.  As compared with the jet and
counterjet, where bins of width 4.7 arcsec and length 3 arcsec contain
less than 100 counts (Fig.~\ref{fig:xrprofile}), PL2 contains $\sim
1450$ counts.  The hardness ratio ($+0.31$) is shown as the filled
circle with its error bar in Figure~\ref{fig:hardness}.  It is
noteworthy that the X-ray hardness is comparable to that of the inner
parts of the resolved jet and counterjet, lending additional weight to
the idea that PL2 represents the bright extension of the jets into the
unresolved core.  In comparison the thermal emission is much softer:
the hardness ratio in the thermal component in the core (using the
same method as for PL2) is $-0.61\pm0.03$, and the galaxy gas in the
region defined in Section~\ref{sec:gasspectra} has a hardness ratio of
$-0.51^{+0.03}_{-0.05}$ ($1\sigma$ errors).  Component PL1 is
much harder than PL2 due to high absorption, with few counts
between 0.3 and 1 keV, and so has a hardness ratio of $+1.0$.
We emphasize that PL2 has an X-ray hardness
ratio that follows the trend seen in the resolved jet.

\section{Discussion}
\label{sec:discusion}

\subsection{Emission mechanism of the jet and counterjet X-rays}
\label{sec:discussion-em}

\begin{figure}
\centering
\includegraphics[width=2.0truein]{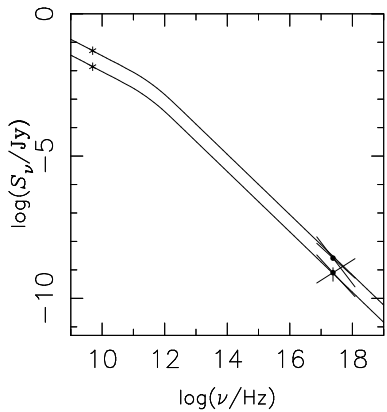}
\caption{
Spectral distribution in the radio and X-ray in the
jet (upper points) and counterjet (lower points) integrated over regions where
resolved X-ray emission is detected.  The curves are broken-power-law
models, $\Delta\alpha=0.5$, with minimum-energy magnetic fields, and
with low-frequency spectra matching a radio spectral index of $\alpha=0.56$.
}
\label{fig:sed}
\end{figure}

Here we consider the radio and X-ray emission from the jet and
counterjet, as extracted from the regions defined in
Section~\ref{sec:xjet}.  A power-law extrapolation to higher
frequencies of the radio emission, with a spectral index of
$\alpha^{5~\rm GHz}_{2.7~\rm GHz} = 0.56$ \citep{wall}, falls above
the X-ray emission, but consistency with the X-ray can be found by
applying a spectral break, $\Delta\alpha$, of similar size in the jet
and counterjet.  Figure~~\ref{fig:sed} shows the result of applying
$\Delta\alpha=0.5$, as expected for the simplest case of
continuous injection where the electrons are losing energy by
synchrotron radiation and inverse Compton scattering.  The
predictions for inverse-Compton X-ray emission are insensitive to the
form of the electron spectrum at high energies, and we have
estimated the minimum-energy magnetic field and the inverse Compton
yield assuming a minimum electron Lorentz factor of 10, no heavy
particles, and a filling factor of unity \citep[see, e.g.,][]{wbrev}.
An inverse-Compton origin for the X-ray emission (from upscattering
starlight, see below) suggests an uncomfortably large (factor
$\sim$~45) departure from minimum energy, in common with many
lower-power jets \citep[e.g.,][]{hard-66b}. An inverse Compton origin
would also predict similar radio and X-ray spectral slopes, which is
allowed by the data for the counterjet but not for the jet.  It is
therefore highly unlikely that the X-ray emission is of inverse
Compton origin, and it is almost certainly produced by synchrotron
radiation.

Using the assumptions above and
the (weak) level of relativistic beaming with parameters given
below (Section~\ref{sec:discussion-im}), we find equal minimum-energy
magnetic fields in the jet and counterjet (2.2~nT).  The spectral
break of $\Delta\alpha=0.5$ in
the synchrotron emission (Fig.~\ref{fig:sed}) is then due to a knee in the electron
spectrum at $\sim 40$~GeV where it breaks from a power-law index of $p = 2.12$
to 3.12.

\citet*{stawarz-starlight} pointed out that starlight can be
the dominant radiation field for inverse Compton scattering of
electrons in kpc-scale jets, and this is the case for \sourcer\ in the
region where X-rays are detected.  From the V-band surface-brightness
profile of \citet{kormendy} we estimate an average starlight energy
density for distances between 10 and 20 arcsec from the nucleus of $9.3 \times
10^{-12}$ J m$^{-3}$, which is to be compared with $4.2 \times 10^{-14}$ J
m$^{-3}$ in the cosmic microwave background and $1.9 \times 10^{-12}$
J m$^{-3}$ in minimum-energy magnetic field.  In our calculation above
of the inverse Compton X-ray contribution from scattering of starlight we have
modelled the starlight as peaking at $10^{14}$~Hz
\citep{stawarz-starlight} and having a diluted black-body spectrum to
match the estimated energy density.

The lifetime of electrons to synchrotron radiation and inverse Compton
losses in the Thomson regime, $\tau$, is given by
 
$$\tau = {3 m_{\rm e} c \over 4 \sigma_{\rm T} u \gamma},\eqno(1)$$

\noindent
where $m_{\rm e}$ is the electron mass, $\sigma_{\rm T}$
is the Thomson cross section, $\gamma$ is the electron Lorentz factor
and $u$ is the total energy density in photons and field.  While it is
a simplification to treat the X-ray jet as a single
emission region, we can check that $\sim
40$~GeV (see above) is a consistent break energy in the electron
spectrum. Electrons of this high an energy are still just within the
Thomson regime for inverse Compton scattering of starlight, and so we
use the combined energy density of $1.1 \times 10^{-11}$ J m$^{-3}$ in
Equation~1 and calculate a minimum-energy loss lifetime of $\sim
10^5$~yr.  This is comparable to the time ($\sim 6 \times 10^4$ yr)
that we estimate jet plasma has been flowing along the X-ray-detected
jet (Section~\ref{sec:discussion-im}), consistent with a model that
had the same population of accelerated
electrons providing both the radio and X-ray emission through
synchrotron radiation.

In Fig.~\ref{fig:xrprofile} it appears as though the X-ray emission
peaks closer to the core (upstream) than the radio.  This is seen in
other synchrotron jets \citep[e.g.][]{hard-66b, dulwich-3c15}.
The X-ray peaks in Fig.~\ref{fig:xrprofile}
must be produced by a series of substructures, since they are broader
than the synchrotron loss distance of the emitting electrons convolved with the
\chandra\ PSF.  By contrast, if synchrotron loss time dictates the
size of radio knots, then the radio peaks would be far broader than
seen in Fig.~\ref{fig:xrprofile}. Some other process, perhaps
expansion, therefore dominates the losses of radio-emitting
electrons in knots.  But, this is clearly a slow process compared with X-ray
loss times, since the radio peaks are wider.  The build up of radio surface
brightness, from the composite structures, with increasing distance
from the core as the plasma flows downstream, is therefore expected to lead to 
offsets in X-ray and radio surface brightness in the sense observed.

Had X-ray variability been detected between the two \chandra\
observations we would have inferred the presence of components of
size a few light years embedded in the jet.  While evidence for small
features embedded in X-ray synchrotron jets is scarce, it is
unmistakable from the flaring of knot HST-1 of M\,87
\citep{harris-m87}.  More recently, variable X-ray emission is claimed
from a light-year scale region of enhanced magnetic field embedded
within a knot in the jet of Pictor\,A \citep{marshall-pica}, a source
that had similar-length \chandra\ exposures to \sourcer\ but is about
30 times further away. We cannot rule out the existence of such
structures within \sourcer\ based on our two exposures
(Section~\ref{sec:xjet}) but, relative to
Pictor\,A, these observations are handicapped by the shorter jet and richer
X-ray emitting atmosphere.

The X-ray to radio flux-density ratios for the jet and counterjet
(Section~\ref{sec:xjet}) correspond to a two-point spectral index of
$\alpha_{\rm rx} \sim 0.95$.  This is somewhat steeper than typical
values of 0.9 or flatter that are reported for one-sided synchrotron
jets at smaller angles to the line of sight \citep[e.g.,][]{worrall01,
hard-66b, worrall-n315}.  The result can be understood, at least
qualitatively, in the context of synchrotron spectra which break and
steepen between the radio and X-ray (e.g., Fig.~\ref{fig:sed})
at a fixed frequency in the source frame: the emission from
two-sided jets is relatively less blueshifted from Doppler
beaming, and so emission at fixed observing frequencies arises from
higher frequencies in the source rest frame, and a steeper
$\alpha_{\rm rx}$ should be measured.

\subsection{Jet components in the X-ray nucleus}
\label{sec:discussion-jetcore}

In Section~\ref{sec:xnucleus} we pointed out that the ratio of X-ray
emission in PL2 to resolved pc-scale radio emission is similar to that
in the jet and counterjet (see also Fig.~\ref{fig:xrratio}). We used
this to support the idea that PL2 is the pc-scale extension of the
kpc-scale X-ray jet and counterjet.  The small component of intrinsic
absorption that is measured is consistent with the pc-scale emission
lying within the few-hundred-pc-scale dust disc detected with \hst\
\citep*{ferrarese}.  The reddening on appropriately small scales is
uncertain due to subtraction of the bright optical nucleus, but
elsewhere within the disc $A_v$ is in the range 0.3--0.6~mag (Ferrarese
1999, private communication), leading to a column density in the range
of 8 to 16 $\times 10^{20}$ cm$^{-2}$ if we adopt the relationship
between reddening and hydrogen column density given by \cite{bh}.  The
result is consistent with the hydrogen column found for PL2 of
$2^{+14}_{-2} \times 10^{20}$~cm$^{-2}$, although errors are large.

The similarity of X-ray to radio ratio between PL2 and the kpc-scale
emission may suggest that PL2 shares the same synchrotron origin as
the resolved jet and counterjet.  However, for VLBI-scale components,
the synchrotron self-Compton (SSC) process is relatively more
important in producing X-rays than for VLA-scale components, and so
must be considered as a possible emission mechanism.  \citet{jones}
give an upper limit of $\sim 1$~mas for the width of the VLBI jet at
10~mas from the core.  The true width would need to be about one
hundredth of this upper limit for the SSC process to produce the X-ray
component PL2 with a minimum-energy magnetic field.  For sizes closer
to the upper limit, synchrotron radiation must be responsible unless
the magnetic energy density is much less than the particle
energy density, so that the VLBI-scale jet is far from minimum
energy.

While it seems most likely that PL2 is a synchrotron
component, the argument can be turned around to infer that the jet
width cannot be less than about 0.01~mas if at minimum energy, else the
X-ray emission would be exceeded by SSC.  Alternatively, the
pc-scale jet could be magnetically dominated.

Radio spectra are typically flat in pc-scale jets, and so the
similarity in X-ray to radio flux ratio with the kpc-scale jet
requires a similar spectral curvature if indeed the X-ray emission
from PL2 is of synchrotron origin.  There are several uncertainties,
but if we adopt a jet width of 1~mas, energy losses from synchrotron
radiation (for a minimum-energy magnetic field) exceed those from the
dominant inverse Compton processes (scattering on AGN light and SSC).
We find a minimum-energy magnetic field of $\sim 3\mu$T, and from
Equation~1 the synchrotron lifetime of electrons of $\sim 1$~GeV
radiating at around the required break frequency of $3 \times 10^{11}$
Hz is roughly 14~yr.  This is similar to the travel time along the
10~mas jet at $0.46~c$ ($\sim 11$~yr).

The X-ray emission mechanisms of PL1 and PL2 cannot be determined on
the basis of their X-ray spectral slopes which, while similar, have
large errors.  However, the ratio of PL1 X-ray to VLBI luminosity is
an order of magnitude larger than for PL2.  This suggests that if PL1
is embedded in the VLBI core and related to the small-scale jet, the
emission mechanism may not be synchrotron.  \cite{hw} made a
statistical argument that if low-power radio galaxies are dimmed,
redshifted, versions of BL\,Lac objects, as suggested by unification
models, and the nuclear X-ray emission is related to the base of the
radio jet, it must be of inverse Compton rather than synchrotron
origin.  The radio core in \sourcer\ is estimated to be $\sim 0.02$
mas in radius for its inverted spectrum to be consistent with partial
self absorption \citep{jones}.  The X-ray emission of PL1 can then
arise from the core via the SSC process in a magnetic field of $\sim
20 \mu$T, which is roughly 1/5 of the minimum-energy value, if beaming
parameters are as for the mas-scale jet.  The alternative is that PL1
is dominated by emission from a radiatively inefficient accretion flow
\citep[e.g.,][]{zezas}

\subsection{Confinement of the kpc-scale jet by the external medium}
\label{sec:discussion-ec}

\begin{figure}
\centering
\includegraphics[width=3.0truein]{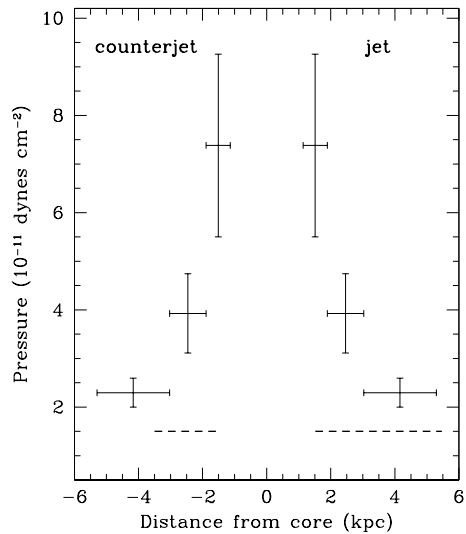}
\caption{ The data points (with 90\% errors) show the pressure of the
external X-ray-emitting atmosphere modelled with spherical symmetry,
from the analysis of \citet{newpaper}.  The dashed lines are the
deprojected average pressures in the counterjet and jet extraction
regions (Section~\ref{sec:xjet}) assuming minimum energy and modelled
as described in Section~\ref{sec:discussion-em}.  Results suggest
that, where resolved X-ray emission is seen, the jet is confined by
the external medium and cannot deviate by a large factor from minimum
energy.  }
\label{fig:pressure}
\end{figure}

The pressures in the X-ray-emitting jet and counterjet plasma
(Fig.~\ref{fig:xrimage}) depend on the
magnetic-field strengths and particle densities.  The minimum
pressures correspond to the minimum-energy condition applied above,
and give values corresponding to the dashed lines in
Figure~\ref{fig:pressure}, where we compare with the external gas
pressure.  The external pressures are from \citet{newpaper} \citep[in
good agreement with][]{hump} and assume spherical symmetry.
The comparison is appropriate if the wedge regions of
lower X-ray counts that appear to
surround the jet and counterjet (Section~\ref{sec:gasspectra}) are in
approximate pressure balance with the cooler galaxy gas and with the
jet and counterjet with
which they are in immediate contact, as suggested by the lack of
evidence for supersonically generated features in the gas
\citep[for more details see][]{newpaper}.
From the pressure comparison
we can see  by how much the jet and counterjet, where X-rays are measured, would
have to exceed their minimum pressures to have
a dynamical effect on the surrounding gas.  Only a small
departure from minimum pressure is required for the gentle
displacement of wedge gas that appears to have taken place. 

Why there is a deficit of counts in wedge-like regions around the jet
(and probably also the counterjet) is unclear.  The regions would
collapse unless there is an unseen
component of pressure.  It is most likely that the pressure is
supplied by old (lobe) electrons that have have lost too much energy
to be radiating at 5~GHz or that radiate with too low a surface
brightness to be detected without the use of shorter-baseline,
lower-frequency, data than used for the image
in Fig.~\ref{fig:xrimage}.  (We note
that the radio data mapped in Fig.~\ref{fig:xrbw}
are significant to $\sim 2.5$ mJy beam$^{-1}$, a fainter flux density
than contoured in the figure, and the lobes then extend much further
back towards the jet.)  However, we would expect to see only
very faint radio emission if any, if the pressure support is from aged
electrons since the electron number density at 40~GeV (the energy corresponding to the
knee in the jet spectrum of Fig.~\ref{fig:sed}) in a minimum-energy magnetic field
would need only to be $\sim 0.008$ per cent of that in the jet to
provide the missing pressure.  This assumes that the electron spectrum
extrapolates down to Lorentz factors of about unity with an
energy-loss spectral index of $p=3.12$.

In the radial analysis of \cite{hump}, the galaxy and group
atmospheres match in density at a radius of about 12~kpc, which is
beyond the positions of the wedges.  Given that the wedges show a
$\sim 40$ per cent deficit in X-ray emissivity, integrated along the
line of sight, as compared with similar regions away from the radio
source, we estimate that $\sim 10$ per cent of the gas mass in the
galaxy has been evacuated by the wedges.  If the volume occupied by
the visible radio structures within the galaxy volume is also taken
into account, this increases to $\sim 20$ per cent of the galaxy gas
mass.  Thus the radio source is having a significant effect in
displacing hot gas within the galaxy, and the work done on the galaxy
gas (a few $\times 10^{50}$~J) will cause appreciable deviations from
hydrostatic equilibrium until the work has been thermalized
over much of the galaxy atmosphere. Only faint evidence for gas
structures is currently seen \citep{newpaper}, suggesting that the
dynamical processes associated with the development of the radio
source were not sudden and violent, and that only weak density and
temperature gradients were created near the cavities containing the
radio lobes. Our estimate of the source age, $\sim 20$~Myr, is too
short for stronger structures to have dissipated.

Over a radio source lifetime of $\sim 20$~Myr, the jet power
required to displace the gas is more than $10^{35}$~W.  This is more
than our estimates of the broad-band radiative power of the pc-scale
and large-scale jets, at $\sim 8 \times 10^{33}$~W and $\sim 3 \times
10^{33}$~W, respectively, and more than the unabsorbed X-ray power
within the absorbed nuclear component of $\sim 1.4 \times 10^{34}$~W.
From this we infer that the jet radiates less than 20 per cent of its
energy.  The powers involved are still relatively modest, however,
requiring only a small fraction of the Eddington luminosity of the
central black hole \citep[$6.5 \times
10^{39}$~W;][]{ferrarese}.

\subsection{Electron lifetime considerations}
\label{sec:discussion-el}

We argue in Section~\ref{sec:discussion-em} that the kpc-scale X-ray
emission is synchrotron radiation.  In the minimum-energy magnetic
field the synchrotron loss time of X-ray-emitting electrons
is of order 800 yr \citep[see
equation~14 of][]{worrall}, which is less than the travel time down
the 4.9~kpc-long X-ray jet (5.5~kpc de-projected for
$\theta=63^\circ$).  In common with other low-power synchrotron jets
resolved with \chandra, this necessitates particle acceleration to
electron energies no less than $\sim 10^{13}$~eV along the jet.

The mechanism of particle acceleration is not well understood.  Some
of it may be triggered by the jet interacting with slow-moving gas
clouds or high-mass stars as, for example, proposed for some knots in
the inner $\sim 1$~kpc of the jet in nearby Cen~A \citep{hard-cena}.
However, Cen~A is sufficiently nearby to allow proper motions of
features in the kpc-scale jet to be measured, and speeds that are a
significant fraction that of light are found at radio
wavelengths in some knots \citep{hard-cena}.  This is inconsistent
with those knots being fixed to something
travelling at stellar or gas speeds.  
The jet and counterjet X-ray emission we resolve in \sourcer\ is
at distances of 1.5~kpc and greater from the core, in the regime where
the Cen~A jet is distinctly less knotty \citep{worr-cena}.
Furthermore, whereas the host galaxy of Cen~A has an inner warped
disc suspected to be the merger remnant of a small gas-rich spiral
galaxy \citep[e.g.,][]{quillen}, there is no reason to believe that
\source\ is particularly rich in cold gas or high-mass stars.

Some jets of radio galaxies that are classed as FRI \citep{fr} like
\sourcer, or are at the low-power end of FRIIs, have features
bright in both radio and X-ray that have been identified with strong
shocks, largely due to the magnetic-field compression evident in
optical polarization data \citep[e.g.,][]{perl-m87, dulwich-3c15,
dulwich-3c346}.  However, strong discrete polarization features do not
dominate the jet emission of the low-power population as a whole
\citep[e.g.,][]{perl-atlas}, and a more distributed mechanism of
particle acceleration is needed.

\citet{worr-cena} found that the knotty component of the inner Cen\,A
jet has a flatter X-ray spectrum closer to the jet axis, and used this
to support the presence 
of stratification, with the strongest shear towards the centre of
the jet so that stronger turbulence and more efficient particle
acceleration might occur there \citep[e.g., as described by][]
{bicknell, eilek, stawarz} and migrate to fill the whole width of the jet as a
consequence of circulation associated with the shear. Stratified
flow is required to explain many of the radio properties
of low-power jets \citep{laing}, and in some cases is
supported by detailed kinematic modelling
\citep[e.g.,][]{laingbridle-31data}.  It may also explain why about 10
per cent of the radio and X-ray emission of the inner jet in NGC\,315
is in a filament of oscillatory appearance, possibly representing a
magnetic strand caught up in the flow \citep{worrall-n315}.  Our
assumption in what follows is that in \sourcer\ most of the particle
acceleration arises as a result of turbulence and dissipation in a
stratified flow.

\subsection{Jet intermittency?}
\label{sec:discussion-im}

In Section~\ref{sec:xjet} we estimated a radio jet-to-counterjet
intensity ratio
of 1.8.  The ratio due to
relativistic boosting is given by $((1 + \beta\cos\theta)/(1 -
\beta\cos\theta))^{(2 + \alpha_{\rm r})}$, and so
for $\theta = 63^\circ$ (Section~\ref{sec:intro}) and $\alpha_{\rm
r}=0.56$ (Section~\ref{sec:discussion-em}), we infer
a flow speed of $\sim 0.25~c$.  This speed is slower than that
measured on VLBI scales (Section~\ref{sec:intro}), but is
consistent with a jet that is decelerating due to entrainment
of material either from the external medium or stellar mass loss
\citep{komissarov}.  Deceleration is also consistent with the
weak evidence above that the X-ray spectrum steepens along the jet,
indicative of less kinetic power being available for particle
acceleration.  However, we note that $0.25c$ is still
highly supersonic with respect to the external medium.

A problem in explaining the jet-counterjet luminosity asymmetry as merely due to
relativistic boosting is that the jet and counterjet
lengths should be the same, whereas the X-ray jet is detected a factor of
1.56 further from the nucleus than the counterjet.  Some of this may
be explained by the fact that the jet X-ray surface brightness is higher,
and so
it can be measured further away.  Excluding the faintest two jet
points in Figure~\ref{fig:xrprofile}, that would not be detections when
de-beamed to the counterjet side, and taking the jet and counterjet as
extending into the nucleus, the jet-to-counterjet length ratio is
1.25.

Since the X-ray emission requires {\it in situ\/} particle
acceleration, a possibility for the length asymmetry is an asymmetry
in particle acceleration.  However, it is notable that the
jet/counterjet length ratio in X-rays matches expectations if the jet
and counterjet properties are matched but changed significantly at
some time in the past, and X-ray emission is detectable only along the
distance that the plasma has travelled ($\sim 6 \times 10^4$ yr if
at 0$.25c$) in the time since this change occurred.  In this case, the
longer light travel time from the counterjet means that we see a
shorter counterjet than jet.  The sidedness ratio in length is given
by $(1 + \beta\cos\theta)/(1 - \beta\cos\theta) \approx 1.25$, in
agreement with our measurement above.

As to the nature of the change in jet property that occurred $\sim 6
\times 10^4$ yr ago, we speculate that it might be an increased shear in
the jet flow, triggering a higher level of
turbulence and dissipation, and corresponding particle acceleration to
X-ray energies, and presumably triggered by the very mechanism that
launches a jet close to the supermassive black hole.  The
intermittency of FRII jet central engines over timescales of $\sim 10^4 -
10^6$ yr has been supported by observational and theoretical
considerations \citep[e.g.,][]{clarke, reybeg, janiuk, stawarz-ep,
siemiginowska, kataoka}.
\sourcer\ would then be an example of an FRI displaying a similar
behaviour, where we require
the intermittency to alter the ability of the jet to
accelerate particles to X-ray energies over kpc scales.

If we wish, more globally, to attribute the length of jet seen in
X-rays to a process originating from the central engine over a
discrete period of time, $\Delta t$, we might expect to see examples of detached
stretches of X-ray jet along a low-power radio jet.  However, there is
growing evidence that X-ray emission from low-power jets is not
observed beyond a zone of rapid deceleration where the jet opening
angle flares significantly, for example 18~kpc from the nucleus in the
case of NGC\,315 \citep{worrall-n315}.  Given the relatively short
time for plasma to reach such a flare point ($\sim 6 \times 10^5$
yr for NGC\,315), the chance of seeing a
detached jet would be small unless $\Delta t$ is typically
much shorter than this.  Also, where a detached jet is seen it
will be interpreted as a knot unless $\Delta t$ is a large fraction
of the time taken to travel to the flare position.  A
possible example warranting further scrutiny is the long jet of
NGC\,6251 \citep{evans-n6251}, where detached stretches of X-ray
emission are seen at (projected) distances of 98--130 and 162--200 kpc
from the nucleus (the angle to the line of sight is uncertain).

\section{Conclusions}
\label{sec:conclusions}

Our \chandra\ observations of \sourcer\ have resolved jet emission out
to 31.7 and 20.3 arcsec in the jet and counterjet, respectively.  The
energy spectral index in the jet is $\alpha = 1.22\pm 0.22$ (90\%
errors), with a consistent, but poorly constrained, value for the
counterjet.  The jet-to-counterjet intensity ratio of $1.6 \pm 0.3$
($1\sigma$ errors) is consistent with that found in the radio, and
suggests only modest slowing of jet plasma to about $0.25c$ from the
speed previously inferred on VLBI scales.  In common with previous
reports for other low-power radio galaxies, we point out that an
interpretation of the X-rays as inverse Compton radiation leads to a
large departure from minimum energy, and we argue in favour of a
synchrotron origin.  The jet and counterjet have remarkably similar
minimum-energy magnetic fields.

As previously reported, the core X-ray spectrum is complex.
However, we find that the X-ray hardness ratio and X-ray-to-radio flux ratio for
the kpc-scale emission extrapolate well to match values measured for
one of the power-law components measured in the core.  We therefore
associate this with an inner, bright, extension of the X-ray
synchrotron jet. There is weak evidence for the
X-ray spectrum of the kpc-scale emission steepening with distance from
the core, as would be consistent with slow jet deceleration and reduced
kinetic power available for particle acceleration.

We observe wedge-like regions deficient in diffuse X-ray emission
surrounding the jets.  The regions would collapse under the weight of
exterior galaxy and group gas unless there is an unseen component of
pressure.  We argue that this can be provided by aged electrons from
the old radio lobes.  Combining the evacuated region with that
occupied by the visible jets, we find that the radio source is
responsible for having displaced as much as 20 per cent of the
mass of the interstellar medium.

The fact that the kpc-scale jet and counterjet are so X-ray bright,
internally similar, and apparently undisturbed by their environment,
makes their observed properties a powerful probe of the source
physics.  In particular, while the X-ray jet-to-counterjet luminosity
ratio is consistent with an origin in relativistic beaming, the length
ratio is not, suggesting instead that it is dictated by
light-travel-time effects.  This is unexpected, since the
synchrotron-loss-lifetime of the X-ray-emitting electrons is short,
and {\it in-situ\/} particle acceleration must be present.  We
speculate that the observations can be reconciled if an event in the
central engine, starting about $6 \times 10^4$ yr ago but more
recent than the birth of the radio source, increased the ability of
the jet to accelerate electrons to energies no less than $\sim
10^{13}$~eV.  This may have arisen through an increased velocity shear
enhancing the ability for particle acceleration to proceed through
turbulence and dissipation.  This suggests that jet X-ray emission in
low-power radio galaxies can be used as a probe of the episodic nature
of AGN activity.  These ideas might be tested by searching for
evidence of other jets in which the X-ray jet length,
characteristically shorter than the radio, cannot be explained merely
by the jet having rapidly decelerated, and for cases where jet X-ray
emission is detached from the core.

\section*{Acknowledgments}

We are grateful to the anonymous referee for suggestions
which improved the clarity and completeness of the paper.
We thank the CXC for its support of \chandra\ observations,
calibrations, data processing and analysis, and the SAO R\&D group for
{\sc DS9} and {\sc funtools}.
%, Alexey Vikhlinin for a copy of his
%{\sc zhtools} software that has been used for part of our analysis.
EOS acknowledges the support of the European Community under the Marie
Curie Research Training Network.  AZ acknowledges partial support from
\chandra\ grant GO8-9094X.  Space Astrophysics in Crete is partly
supported by EU FP7 {\it Capacities\/} GA No206469.
This work has used data
from the \vla.  NRAO is a facility of the National Science Foundation
operated under cooperative agreement by Associated Universities, Inc.

% If 3 authors, the first time cite it as \citep*{}

\end{document}